\documentclass[%
reprint,
showpacs,
 amsmath,amssymb,
 aps,
 prl
]{revtex4-1}

\newcommand{\beq}{\begin{equation}}
\newcommand{\eeq}{\end{equation}}

\newcommand{\pdhfrac}[2]{\mathchoice{\frac{#1}{#2}}{#1/#2}{#1/#2}{#1/#2}}

\newcommand{\pd}[2]{\pdhfrac{{\partial}#1}{{\partial}#2}}

\usepackage{subfigure}
\usepackage{color}
\usepackage{graphicx}
\usepackage{dcolumn}
\usepackage{bm}

\begin{document}


\title{The Role of Grain Boundaries under Long-Time Radiation}

\author{Yichao Zhu}
\author{Jing Luo}
\author{Xu Guo}%
 \email{guoxu@dlut.edu.cn}
\affiliation{%
 State Key Laboratory of Structural Analysis for Industrial Equipment, Department of Engineering Mechanics, Dalian University of Technology, Dalian, Liaoning, 116023, China
}%
\affiliation{International Research Center for Computational Mechanics, Dalian University of Technology}


\author{Yang Xiang}
\affiliation{
 Department of Mathematics, The Hong Kong University of Science and Technology, Clear Water Bay, Kowloon, Hong Kong
}%

\author{Stephen Jonathan Chapman}
\affiliation{%
 Mathematical Institute, University of Oxford, Andrew Wiles Building, Radcliffe Observatory Quarter, Woodstock Road
Oxford, OX2 6GG, U.K.
}%


\date{\today}

\begin{abstract}
  Materials containing a high proportion of grain boundaries offer significant potential for the development of radiation-resistent structural materials. However, a proper understanding of the connection between the radiation-induced microstructural behaviour of grain boundary and its impact at long natural time scales is still missing. In this letter, point defect absorption at interfaces is summarised by a jump Robin-type condition at a coarse-grained level, wherein the role of interface microstructure is effectively taken into account. Then a concise formula linking the sink strength of a polycrystalline aggregate with its grain size is introduced, and is well compared with experimental observation. Based on the derived model, a coarse-grained formulation incorporating the coupled evolution of grain boundaries and point defects is proposed, so as to underpin the study of long-time morphological evolution of grains induced by irradiation. Our simulation results suggest that the presence of point defect sources within a grain further accelerates its shrinking process, and radiation tends to trigger the extension of twin boundary sections.
  \footnote{Accepted by $Physical$ $Review$ $Letters$ on May 4th, 2018}
\end{abstract}

\maketitle

It is widely agreed that interfaces, such as grain boundaries (GBs), can act as sinks to crystalline point defects (PDs), i.e., interstitial atoms and vacancies \cite{Watanabe_JNuclMater2000, Ackland_Science2010, Beyerlin_MaterToday2013, ElAtwani_MicroscMicroanal2015}. Thus enhancing the GB fraction in materials seems a promising way to develop materials serving in radiation environments, where point defects are prevalent \cite{Zinkle_ActaMater2013, Was_book2017}. At present, the most tractable method for capturing the long-term impact of GB interaction with radiation-induced PDs on materials mechanical properties, is to examine the underlying mechanisms on a continuum scale, where PDs are measured by their concentration distributions. A conventional treatment assumes GBs to be perfect PD sinks, i.e., the PD concentration is maintained at a constant equilibrium value in the vicinity of a GB \cite{Was_book2017}. Nevertheless, this assumption contradicts many experimental observations and atomistic simulation results showing that interfaces are partial PD sinks \cite{Siegel_ActaMetal1980}, and the sink efficiency is highly dependent on GB's microstructural details \cite{Bai_Science2010, ZhangYF_JNuclMater2012, Khater_ActaMater2012, Rajabzadeh_PRL2013, Shao_SciRep2013, Beyerlein_ProgMaterSci2015, Field_ActaMater2015, Ardell_COSSMS2016}. A starting point to address this issue is to consider the PD sink efficiency of low-angle tilt grain boundaries (LATGBs), which can be envisaged as comprising single/multiple sets of edge dislocations \cite{Hirth_book1982}. Then the PD-LATGB interactions should be mediated by the climb motion of grain boundary dislocations (GBDs) \cite{Arzt_ActaMetall1983, Bachurin_ActaMater2012}. In this direction, a Robin-type boundary condition (BC) was recently proposed under a highly symmetric set-up \cite{GuYJ_JMPS2017}. From an engineering perspective, the overall PD sink strength of polycrystalline aggregates is more related to the design of radiation-tolerant structural materials \cite{Beyerlein_ProgMaterSci2015}. All the aforementioned treatments of PD-GB interactions, however, can not be adopted to serve this goal, because they do not capture the fraction of PDs penetrating to neighbouring grains. In this letter, by summarising the underlying PD-GBD interaction mechanisms, we derive a jump Robin-type BC which formulates interfaces as partial sinks to PDs. Fueled by the proposed formulation, a concise formula linking the PD sink strength of polycrystals with grain size is derived. A model incorporating the coupled evolution of PDs and LATGBs is further introduced to underpin the investigation of the long-time radiation-induced behaviour of GBs.

\begin{figure}[ht]
  \includegraphics[width=.48\textwidth]{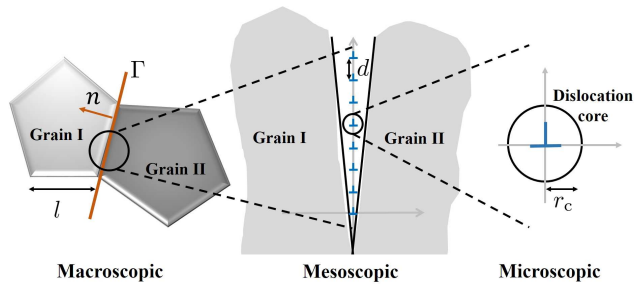}
  \caption{The three regimes characterised by average grain size $l$, the GBD spacing $d$ and the magnitude of the Burgers vector $b$, respectively.\label{fig_3scales} }
\end{figure}

We consider a simplified two-dimensional polycrystalline configuration as shown in the leftmost panel of Fig.~\ref{fig_3scales}. The PD distribution is measured by its (non-dimensional) concentration fraction denoted by $c(x,y,t)$. In grain interiors, it satisfies \cite{Was_book2017, Hirth_book1982}
\beq\label{eqn_c}
\pd{c}{t}= -\nabla\cdot\textbf{J} + S,
\eeq
where ``$\nabla$'' denotes the spatial gradient (with respect to $x$ and $y$); $S$ measures the PD generation rate due to irradiation. The PD flux $\textbf{J}$ appearing in Eq.~\eqref{eqn_c} is given by \cite{Hirth_book1982}
\beq \label{flux}
\textbf{J} = -D\nabla c + D\frac{c\Delta\Omega}{k_\text{B}T}\nabla p,
\eeq
where $D$ is the diffusion coefficient; $p$ is the hydrostatic pressure related to the stress field $\boldsymbol{\sigma}$ by $p=-\text{tr}(\boldsymbol{\sigma})/3$; $k_{\text{B}}$ is the Boltzmann constant; $T$ is the temperature; and $\Delta\Omega$ denotes the PD relaxation volume. The first term in Eq.~\eqref{flux} is just from the first Fick's law, while the second term arises from the work done by the hydrostatic pressure as a PD is absorbed. In 2D, the crystalline mismatch between neighbouring grains is measured by a small misorientation angle $\theta$. From a microstructural viewpoint, such a mismatch is accommodated by $M$ sets of GBDs associated with Burgers vector $\{\textbf{b}^i\}_{i=1}^M$, respectively, and the PD-LATGB reaction can be envisaged as being carried out by the PD-GBD interactions. Here we assume that a GBD acts as a perfect sink of PDs, i.e., the PD concentration fraction within the core region of a GBD is maintained at the constant thermal equilibrium value $c_{\text{e}}$ \cite{Hirth_book1982, GuYJ_JMPS2015}.

The formulation above is established on a length scale short enough to resolve dislocation cores. However, when the long-time radiation-induced behaviour is concerned, a coarse-grained formulation is preferred, and the key question is how to effectively upscale the underlying physics. To better elucidate our coarse-graining strategy, we first consider a simple case of symmetric LATGBs (or twin boundaries), as shown in Fig.~\ref{fig_3scales}.
Three length scale parameters are associated with the problem: the magnitude of the Burgers vector $b$, the GBD spacing $d=b/\theta$ and the average grain size $l$, with $b \ll d \ll l$. Our aim is to build models at the macroscopic level (characterised by $l$), where the LATGB is treated as a curve denoted by $\Gamma$ as shown in the leftmost panel of Fig.~\ref{fig_3scales}. When zoomed into a mesoscopic level (characterised by $d$), GBDs become observable as discrete point PD sinks. When further zoomed into the vicinity of an individual GBD (characterised by $b$), the PD flux should be predominantly driven by two factors: the concentration difference between the core region ($c=c_{\text{e}}$) and the inter-core region of the GB, and the almost-singular hydrostatic pressure field caused by the GBD \cite{Ham_JAP1959, Geslin_PRL2015}. Note that the aforementioned micro/mesoscale processes take place much faster than macroscopic PD evolution, because the underlying PD atoms only need to travel short distances ($\sim b$ or $\sim d$). Such a contrast in evolution speed enables formulating fine-scale dynamics as quasi-steady processes at the coarse-grained level \cite{Chapman_SIAP2016, ZhuYC_JMPS2016}. This idea is carried out more formally by using a three-scale asymptotic analysis (see supplemental material \cite{Supp_PRL2018} for details), which mirrors that in the study of the shielding effect of the Faraday cages \cite{Chapman_SIAMRev2015}, and the PD-LATGB interaction is found to be governed by a jump Robin-type BC at the macroscopic level:
\beq \label{jump_condition}
\left[\pd{c}{\textbf{n}}\right]_{-}^{+} = \omega_0 \cdot (c-c_{\text{e}}) \text{ on }\Gamma
\eeq
with
\beq \label{sink_efficiency}
\omega_0 = \frac{2\pi \theta}{\ln{\frac{2}{\pi \theta}} - h}\cdot\frac1{b},
\eeq
where $\left[\cdot\right]_-^+$ measures the directional quantity jump across the LATGB (so with reference to the leftmost panel of Fig.~\ref{fig_3scales}, $[\frac{\partial c}{\partial \textbf{n}}]_-^+$ equals the derivative of $c$ along $\textbf{n}$ evaluated on the grain I side subtracted by that on the grain II side); $h$ is a material parameter whose expression is given by Eq.~(1.30) in the supplemental material \cite{Supp_PRL2018}.

With reference to Eq.~\eqref{flux} and the fact that both $c$ and $p$ are continuous across the LATGB, the derivative jump in Eq.~\eqref{jump_condition} quantifies the difference in PD flux across the LATGB, which further equals the PD number absorbed per second per area. Eq.~\eqref{jump_condition} also suggests that the PD absorption rate should scale with the excess PD concentration fraction in GB vicinity. Hence the parameter $\omega_0$ can be employed as a measure to the PD sink efficiency of the LATGB. The limit $\omega_0\rightarrow\infty$ corresponds to a perfect sink, i.e. $\left.c\right|_{\Gamma}=c_{\text{e}}$. More importantly $\omega_0$ carries microstructural information about the LATGB, such as the Burgers vector of its constituting dislocations and the misorientation angle, etc. Eq.~\eqref{sink_efficiency} reads that the sink efficiency $\omega_0$ decreases with the misorientation angle $\theta$, which agrees with experimental observation \cite{Siegel_ActaMetal1980}.

Although Eq.~\eqref{jump_condition} is derived in the context of twin boundaries, its mathematical structure of jump Robin BC is still applicable for describing PD's interaction with more complicated three-dimensional (3D) GBD microstructures \cite{KangK_JAP2012, Hirth_ProgMaterSci2013, WangJ_JMaterRes2013, Shao_SciRep2013}, because it reveals the key underlying physics: interfaces are partial PD sinks whose strength is determined by their microstructures; the local PD absorption rate which is measured by the PD flux difference across interfaces should increase with the excess PD concentration in GB vicinity. For situations where 3D interfaces are involved, the corresponding PD sink efficiency coefficient $\omega_0$ needs to be modified with respect to GB microstructures, e.g., through summarising the underlying atomistic simulation results.

\begin{figure}[!ht]
  \includegraphics[width=.35\textwidth]{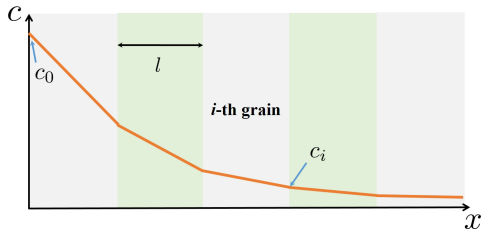}
  \caption{A polycrystalline specimen consisting of laminar grains of width $l$ is considered. Under this set-up, the PD concentration fraction takes a piecewisely linear profile at steady states. \label{fig_layered_grains} }
\end{figure}

\begin{figure}
  \includegraphics[width=.48\textwidth]{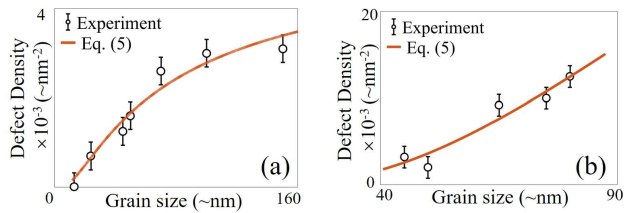}
  \caption{Comparison with experimental data \cite{Rose_NIMPRB1997}. The defect density in Ref.~\cite{Rose_NIMPRB1997} is measured in cluster/nm$^2$. (a) ZrO$_2$ (4MeV, 1$\times10^{16}$Kr/nm$^{2}$), parameter values: $c_0=5.06$nm$^{-2}$, $\omega_0=2.16$nm$^{-1}$, $L/k=13$nm; (b) Pd (240keV, 2$\times10^{16}$Kr/nm$^{2}$), $c_0=198.5$nm$^{-2}$, $\omega_0=1.71$nm$^{-1}$, $L/k=45$nm;\label{fig_comp_exp}}
\end{figure}

Based on the proposed jump Robin BC~\eqref{jump_condition}, we now investigate the role of grain size in determining the PD sink strength of polycrystalline aggregates. To this end, we consider a polycrystalline specimen consisting of laminar grains of thickness $l$ as shown in Fig.~\ref{fig_layered_grains}.
The steady state of the system is considered for theoretical convenience, i.e., $c=c(x)$. Moreover, we suppose PDs are  generated from the left side of the specimen. Thus Eq.~\eqref{eqn_c} becomes $\frac{\mathrm{d}^2 c}{\mathrm{d} x^2}=0$. With Eq.~\eqref{jump_condition}, $c$ is found to be piecewisely linear in space as shown in Fig.~\ref{fig_layered_grains}. If we use $c_i$ to denote the PD concentration fraction measured at the $i$-th GB, then $c_i$ should satisfy a recursive relation (given by Eq.~(2.2) in the supplemental material \cite{Supp_PRL2018}). The recursive equation is further shown \cite{Supp_PRL2018} controlled by a non-dimensional number $\omega_0l$ with $\omega_0$ the average sink efficiency of all GBs in the polycrystal. Practically, $\omega_0l$ takes large values \cite{Supp_PRL2018}. Hence we derive \cite{Supp_PRL2018}:
\beq \label{PD_concentration_overall}
c_{\text{a}} = c_0\left(\frac1{\omega_0l}\right)^{\frac{L}{kl}},
\eeq
where $c_{\text{a}}$ measures the average PD concentration fraction within the polycrystal; $c_0$ measures the PD concentration fraction surrounding irradiation sites, which is further related to irradiation dose; $L$ quantifies the dimension of the polycrystal; $k$ is an adjusting parameter as the simplified laminar-layered model is applied for 3D configurations. Here we let the equilibrium concentration $c_{\text{e}}=0$. A comparison with experimental data \cite{Rose_NIMPRB1997} in Fig.~\ref{fig_comp_exp} shows that among numerous mutually-convoluted physical processes taking place on fine scales, Eq.~\eqref{PD_concentration_overall} well captures the trend in PD sink strength against grain size.
Note that the defect density in Ref.~\cite{Rose_NIMPRB1997} is reported in (defect) cluster/nm$^2$. Both $c_{\text{a}}$ and $c_0$ in Eq.~\eqref{PD_concentration_overall} should be adjusted in unit of nm$^{-2}$, and curve fitting is needed when compared with experiments. However, Eq.~\eqref{PD_concentration_overall} still sees its physical significance due to the following reasons. First, parameters appearing in Eq.~\eqref{PD_concentration_overall} ($c_0$, $\omega_0$ and $L/k$) have their physical meanings, and their fitted values all fall in their corresponding physical ranges (See Fig.~\ref{fig_comp_exp}). Moreover, Eq.~\eqref{PD_concentration_overall} appropriately rationalises several issues addressed in Ref.~\cite{Rose_NIMPRB1997}, such as the slope changes in $c_{\text{a}}$ against $l$ and the existence of a possible saturation value in $c_{\text{a}}$ as grain size becomes large.

Eq.~\eqref{PD_concentration_overall} enjoys a concise form, and it provides an effective means for analysing the PD sink strength of polycrystals. For instance, a practically meaningful question is what value the minimum dimension a radiation-resistent material should take in order to achieve a certain degree of PD shielding capability, which can be quantified by $\alpha=\ln(c_0/c_{\text{a}})$. By rearranging Eq.~\eqref{PD_concentration_overall}, the minimum dimension is found to satisfy
\beq \label{effective_thickness}
L = k\cdot \frac{\alpha l}{\log\left(\omega_0l\right)},
\eeq
which helps address the engineering significance of the present study.

\begin{figure*}
  \includegraphics[width=.9\textwidth]{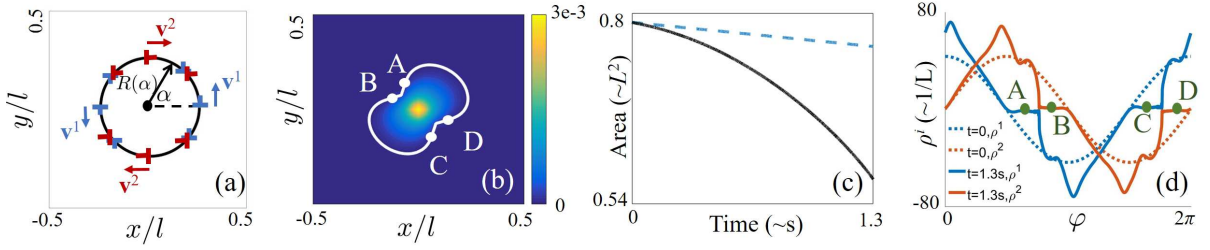}
  \caption{Coupled evolution of PD concentration fraction and LATGB profile. (a) Initial configuration, a cylinderical grain of radius $l/4$ with $l=800$nm. GB sections equipped with different dislocation characters move asymmetrically. (b) GB shape at $t=1.3$s and the distribution of PD concentration fraction. (c) The evolution of the area enclosed by the LATGB with and without PD sources. (d) Profiles of GBD densities on the LATGB. \label{fig_example}}
\end{figure*}
Accumulated evidence has also indicated that radiation may result in morphological changes to material microstructures \cite{Beyerlin_MaterToday2013}. With Eq.~\eqref{jump_condition}, the study in this area may also become feasible. To this end, we consider, as shown in Fig.~\ref{fig_example}(a), a cylindrical grain with its LATGB formed by two GBD species:  $\textbf{b}^1 = (b,0)$ and $\textbf{b}^2 = (0,b)$. Note that the GBD configuration is also represented at a coarse-grained level, i.e., the GBDs are described by their density distribution. This is accomplished by assigning to the $i$-th GBD set a dislocation density potential function (DDPF) $\eta^i(x,y,t)$, such that the $i$-th GBD density (number per unit length in this case) is given by $\rho^i = \frac{\partial \eta^i}{\partial s}$ with $s$ the GB arclength. Note that $\rho^i$ takes negative values when the actual Burgers vector of the GBDs is opposite to the original set-up. Hence the evolution of GBD ensembles is formulated by \cite{ZhuXH_JMPS2014}
\beq \label{DDPF_evolution}
\pd{\eta^i}{t} + \textbf{v}^i \cdot \nabla\eta^i = 0,
\eeq
where $\textbf{v}^i$ is the velocity field associated with the $i$-th GBD set. Here we consider the case where the diffusional climb of GBDs is predominant \cite{Arzt_ActaMetall1983}, i.e. $|\textbf{v}^i|=|v_{\text{c}}^i|$ with $v_{\text{c}}^i$ denoting the climb component. Here a positive $v_{\text{c}}^i$ corresponds to the case where a PD is absorbed. Based on the fact that the number of PDs absorbed/released by an LATGB section should equal the overall lattice spacings climbed by all GBDs on that section, the GBD climb speed is thus related to the PD absorption rate by
\beq \label{conservation_absorption}
\sum_{i=1}^M \rho^i v_{\text{c}}^i b^i= D\omega_0\cdot(c-c_{\text{e}}),
\eeq
whose derivation is detailed in the supplemental material \cite{Supp_PRL2018}. Finally, since the GBD motion should be consistent with the LATGB's along the GB normal, we require that the GBD velocity components along the GB normal be identical, i.e. $v_n = \textbf{v}^i\cdot\textbf{n}$ stays the same for all $i=1$, $\cdots$, $M$. Thus the instantaneous LATGB migration speed satisfies
\beq \label{speed_GB}
v_{\text{gb}} = v_n - \gamma\kappa,
\eeq
where $\kappa$ is the signed GB curvature; $\gamma$ is a GB energy parameter. The second term of Eq.~\eqref{speed_GB} is due to the GB capillary effect \cite{Sutton_book1995}.

Therefore, a coarse-grained equation system governing the underlying PD-LATGB evolution is derived, and a simulation example is presented in Fig.~\ref{fig_example}. Here the LATGB is represented in a polar coordinate system by $R(\varphi,t)$ as shown in Fig.~\ref{fig_example}(a) \cite{ZhangLC_arXiv2017}. For simplicity we let $p$ in Eq.~\eqref{flux} vanish. The simulation starts with a grain of cross-sectional radius $200$nm, at whose centre a PD source $S=2\times10^{-3}D\delta{(x)}\delta{(y)}$ is located. At $t=0$, the misorientation angle $\theta=2.5^{\circ}$. Note that the initial GBD density is fully determined by $\textbf{b}^1$, $\textbf{b}^2$, $\theta$ and the GB geometry, and in Fig.~\ref{fig_example}(a) the corresponding GBD characters at $\varphi=k\pi/4$ with $k=0$, $\cdots$, $7$ are schematically shown. Other parameters are: $G=40$GPa; $\nu=0.3$; $T=400K$; $\Delta \Omega=4.2\mathring{\text{A}}^3$; $r_{\text{c}}=0.1$nm; $b=0.3$nm; $D=3\times10^{5}$nm$^2$/s \cite{Keralavarma_PRL2012}; $\gamma=0.5$nm$^2$/s. The GB profile at $t=1.3$s is shown by the white curve in Fig.~\ref{fig_example}(b), and the corresponding PD concentration fraction is shown by the colour map. It is observed that the LATGB sections initially lying in the second and fourth quadrants move towards each other, while the other parts migrate apart. Such asymmetry in GB motion results from the fact that GBDs of different characters climb along different directions. The climbing directions of GBDs at selected points are marked in Fig.~\ref{fig_example}(a). Noted that this asymmetric movement of GB can not be captured by the widely used curvature driven law for GB motion. Moreover, as shown in Fig.~\ref{fig_example}(c), compared to the purely curvature driven GB motion, the presence of a PD source inside a grain further accelerates its shrinking process. This is because the PD-LATGB interactions become stronger for the GB sections that move towards the source, as the PD concentration fraction is higher there. In Fig.~\ref{fig_example}(d) the profiles of the densities of different GBD species are plotted against $\varphi$, the angular variable in the polar coordinate system. It shows that as irradiation proceeds, the GB sections wherein the density of one GBD species vanish tends to extend (as marked A-D correspondingly in both Fig~\ref{fig_example}(b) and (d)). This means that radiation is prone to dissociating different GBD species, and twin boundary regions emerge as a result.

The simulation example demonstrates that the derived formulation can be adopted to study the long-time radiation-induced behaviour of GBs, while the underlying microstructure evolution is effectively taken into account. Since the continuum representation of GBD networks with DDPFs also work for 3D configurations \cite{ZhuXH_JMPS2014}, more 3D microscale mechanisms, such as PD pipe diffusion \cite{NiuXH_JMPS2017}, should be incorporated into the proposed framework with limited modification.

In summary, interfaces are modelled as partial PD sinks by a jump Robin BC~\eqref{jump_condition}, which effectively relates interface PD sink efficiency to its microstructures. Then the role of grain size in determining the sink strength of polycrystalline aggregates is examined with a concise formula (Eq.~\eqref{PD_concentration_overall}). Moreover, a coarse-grained framework incorporating PD-GB evolution is also proposed, so as to underpin the study of the radiation-induced morphological evolution of GBs on a time scale that is more closer to natural time scales. From a methodological perspective, the presented upscaling techniques may become a paradigm for exploring the continuum formalism involving interactions between various types of defects for extended applications \cite{Mordehai_PhilMag2008, Geers_JMPS2014, ZhuYC_JMPS2017a,Rovelli_JMPS2017}.

We would like to thank the (anonymous) reviewers for their valuable comments. The work of Y.C.Z., J.L. and X.G. was partly supported by the Natural Science Foundation of China (11772076, 11732004), the Fundamental Research Funds for the Central Universities (DUT16RC(3)091). Y.X. was partly supported by the Hong Kong Research Grants Council General Research Fund 16302115.

\bibliography{mybib}

\begin{thebibliography}{39}%
\makeatletter
\providecommand \@ifxundefined [1]{%
 \@ifx{#1\undefined}
}%
\providecommand \@ifnum [1]{%
 \ifnum #1\expandafter \@firstoftwo
 \else \expandafter \@secondoftwo
 \fi
}%
\providecommand \@ifx [1]{%
 \ifx #1\expandafter \@firstoftwo
 \else \expandafter \@secondoftwo
 \fi
}%
\providecommand \natexlab [1]{#1}%
\providecommand \enquote  [1]{``#1''}%
\providecommand \bibnamefont  [1]{#1}%
\providecommand \bibfnamefont [1]{#1}%
\providecommand \citenamefont [1]{#1}%
\providecommand \href@noop [0]{\@secondoftwo}%
\providecommand \href [0]{\begingroup \@sanitize@url \@href}%
\providecommand \@href[1]{\@@startlink{#1}\@@href}%
\providecommand \@@href[1]{\endgroup#1\@@endlink}%
\providecommand \@sanitize@url [0]{\catcode `\\12\catcode `\$12\catcode
  `\&12\catcode `\#12\catcode `\^12\catcode `\_12\catcode `\%12\relax}%
\providecommand \@@startlink[1]{}%
\providecommand \@@endlink[0]{}%
\providecommand \url  [0]{\begingroup\@sanitize@url \@url }%
\providecommand \@url [1]{\endgroup\@href {#1}{\urlprefix }}%
\providecommand \urlprefix  [0]{URL }%
\providecommand \Eprint [0]{\href }%
\providecommand \doibase [0]{http://dx.doi.org/}%
\providecommand \selectlanguage [0]{\@gobble}%
\providecommand \bibinfo  [0]{\@secondoftwo}%
\providecommand \bibfield  [0]{\@secondoftwo}%
\providecommand \translation [1]{[#1]}%
\providecommand \BibitemOpen [0]{}%
\providecommand \bibitemStop [0]{}%
\providecommand \bibitemNoStop [0]{.\EOS\space}%
\providecommand \EOS [0]{\spacefactor3000\relax}%
\providecommand \BibitemShut  [1]{\csname bibitem#1\endcsname}%
\let\auto@bib@innerbib\@empty
\bibitem [{\citenamefont {Watanabe}\ \emph {et~al.}(2000)\citenamefont
  {Watanabe}, \citenamefont {Takamatsu}, \citenamefont {Sakaguchi},\ and\
  \citenamefont {Takahashi}}]{Watanabe_JNuclMater2000}%
  \BibitemOpen
  \bibfield  {author} {\bibinfo {author} {\bibfnamefont {S.}~\bibnamefont
  {Watanabe}}, \bibinfo {author} {\bibfnamefont {Y.}~\bibnamefont {Takamatsu}},
  \bibinfo {author} {\bibfnamefont {N.}~\bibnamefont {Sakaguchi}}, \ and\
  \bibinfo {author} {\bibfnamefont {H.}~\bibnamefont {Takahashi}},\ }\href@noop
  {} {\bibfield  {journal} {\bibinfo  {journal} {J. Nucl. Mater.}\ }\textbf
  {\bibinfo {volume} {283-287}},\ \bibinfo {pages} {152} (\bibinfo {year}
  {2000})}\BibitemShut {NoStop}%
\bibitem [{\citenamefont {Ackland}(2010)}]{Ackland_Science2010}%
  \BibitemOpen
  \bibfield  {author} {\bibinfo {author} {\bibfnamefont {G.}~\bibnamefont
  {Ackland}},\ }\href@noop {} {\bibfield  {journal} {\bibinfo  {journal}
  {Science}\ }\textbf {\bibinfo {volume} {327}},\ \bibinfo {pages} {1587}
  (\bibinfo {year} {2010})}\BibitemShut {NoStop}%
\bibitem [{\citenamefont {Beyerlein}\ \emph {et~al.}(2013)\citenamefont
  {Beyerlein}, \citenamefont {Caro}, \citenamefont {Demkowicz}, \citenamefont
  {Mara}, \citenamefont {Misra},\ and\ \citenamefont
  {Uberuaga}}]{Beyerlin_MaterToday2013}%
  \BibitemOpen
  \bibfield  {author} {\bibinfo {author} {\bibfnamefont {I.~J.}\ \bibnamefont
  {Beyerlein}}, \bibinfo {author} {\bibfnamefont {A.}~\bibnamefont {Caro}},
  \bibinfo {author} {\bibfnamefont {M.~J.}\ \bibnamefont {Demkowicz}}, \bibinfo
  {author} {\bibfnamefont {N.~A.}\ \bibnamefont {Mara}}, \bibinfo {author}
  {\bibfnamefont {A.}~\bibnamefont {Misra}}, \ and\ \bibinfo {author}
  {\bibfnamefont {B.~P.}\ \bibnamefont {Uberuaga}},\ }\href@noop {} {\bibfield
  {journal} {\bibinfo  {journal} {Mater. Today}\ }\textbf {\bibinfo {volume}
  {16}},\ \bibinfo {pages} {443} (\bibinfo {year} {2013})}\BibitemShut
  {NoStop}%
\bibitem [{\citenamefont {El-Atwani}\ \emph {et~al.}(2015)\citenamefont
  {El-Atwani}, \citenamefont {Hattar}, \citenamefont {Vetterick},\ and\
  \citenamefont {Taheri}}]{ElAtwani_MicroscMicroanal2015}%
  \BibitemOpen
  \bibfield  {author} {\bibinfo {author} {\bibfnamefont {O.}~\bibnamefont
  {El-Atwani}}, \bibinfo {author} {\bibfnamefont {K.}~\bibnamefont {Hattar}},
  \bibinfo {author} {\bibfnamefont {G.}~\bibnamefont {Vetterick}}, \ and\
  \bibinfo {author} {\bibfnamefont {M.~L.}\ \bibnamefont {Taheri}},\
  }\href@noop {} {\bibfield  {journal} {\bibinfo  {journal} {Microsc
  Microanal}\ }\textbf {\bibinfo {volume} {21}},\ \bibinfo {pages} {0200}
  (\bibinfo {year} {2015})}\BibitemShut {NoStop}%
\bibitem [{\citenamefont {Zinkle}\ and\ \citenamefont
  {Was}(2013)}]{Zinkle_ActaMater2013}%
  \BibitemOpen
  \bibfield  {author} {\bibinfo {author} {\bibfnamefont {S.~J.}\ \bibnamefont
  {Zinkle}}\ and\ \bibinfo {author} {\bibfnamefont {G.~S.}\ \bibnamefont
  {Was}},\ }\href@noop {} {\bibfield  {journal} {\bibinfo  {journal} {Acta
  Mater.}\ }\textbf {\bibinfo {volume} {61}},\ \bibinfo {pages} {735} (\bibinfo
  {year} {2013})}\BibitemShut {NoStop}%
\bibitem [{\citenamefont {Was}(2017)}]{Was_book2017}%
  \BibitemOpen
  \bibfield  {author} {\bibinfo {author} {\bibfnamefont {G.~S.}\ \bibnamefont
  {Was}},\ }\href@noop {} {\emph {\bibinfo {title} {Fundamentals of Radiation
  Materials Science}}},\ \bibinfo {edition} {2nd}\ ed.\ (\bibinfo  {publisher}
  {Springer},\ \bibinfo {address} {New York},\ \bibinfo {year}
  {2017})\BibitemShut {NoStop}%
\bibitem [{\citenamefont {Siegel}\ \emph {et~al.}(1980)\citenamefont {Siegel},
  \citenamefont {Chang},\ and\ \citenamefont
  {Balluffi}}]{Siegel_ActaMetal1980}%
  \BibitemOpen
  \bibfield  {author} {\bibinfo {author} {\bibfnamefont {R.~W.}\ \bibnamefont
  {Siegel}}, \bibinfo {author} {\bibfnamefont {S.~M.}\ \bibnamefont {Chang}}, \
  and\ \bibinfo {author} {\bibfnamefont {R.~W.}\ \bibnamefont {Balluffi}},\
  }\href@noop {} {\bibfield  {journal} {\bibinfo  {journal} {Acta Metall.}\
  }\textbf {\bibinfo {volume} {28}},\ \bibinfo {pages} {249} (\bibinfo {year}
  {1980})}\BibitemShut {NoStop}%
\bibitem [{\citenamefont {Bai}\ \emph {et~al.}(2010)\citenamefont {Bai},
  \citenamefont {Voter}, \citenamefont {Hoagland}, \citenamefont {Nastasi},\
  and\ \citenamefont {Uberuaga}}]{Bai_Science2010}%
  \BibitemOpen
  \bibfield  {author} {\bibinfo {author} {\bibfnamefont {X.~M.}\ \bibnamefont
  {Bai}}, \bibinfo {author} {\bibfnamefont {A.~F.}\ \bibnamefont {Voter}},
  \bibinfo {author} {\bibfnamefont {R.~G.}\ \bibnamefont {Hoagland}}, \bibinfo
  {author} {\bibfnamefont {M.}~\bibnamefont {Nastasi}}, \ and\ \bibinfo
  {author} {\bibfnamefont {B.~P.}\ \bibnamefont {Uberuaga}},\ }\href@noop {}
  {\bibfield  {journal} {\bibinfo  {journal} {Science}\ }\textbf {\bibinfo
  {volume} {327}},\ \bibinfo {pages} {1631} (\bibinfo {year}
  {2010})}\BibitemShut {NoStop}%
\bibitem [{\citenamefont {Zhang}\ \emph {et~al.}(2012)\citenamefont {Zhang},
  \citenamefont {Huang}, \citenamefont {Millett}, \citenamefont {Tonks},
  \citenamefont {Wolf},\ and\ \citenamefont
  {Phillpot}}]{ZhangYF_JNuclMater2012}%
  \BibitemOpen
  \bibfield  {author} {\bibinfo {author} {\bibfnamefont {Y.~F.}\ \bibnamefont
  {Zhang}}, \bibinfo {author} {\bibfnamefont {H.~C.}\ \bibnamefont {Huang}},
  \bibinfo {author} {\bibfnamefont {P.~C.}\ \bibnamefont {Millett}}, \bibinfo
  {author} {\bibfnamefont {M.}~\bibnamefont {Tonks}}, \bibinfo {author}
  {\bibfnamefont {D.}~\bibnamefont {Wolf}}, \ and\ \bibinfo {author}
  {\bibfnamefont {S.~R.}\ \bibnamefont {Phillpot}},\ }\href@noop {} {\bibfield
  {journal} {\bibinfo  {journal} {J. Nucl. Mater.}\ }\textbf {\bibinfo {volume}
  {422}},\ \bibinfo {pages} {69} (\bibinfo {year} {2012})}\BibitemShut
  {NoStop}%
\bibitem [{\citenamefont {Khater}\ \emph {et~al.}(2012)\citenamefont {Khater},
  \citenamefont {Serra}, \citenamefont {Pond},\ and\ \citenamefont
  {Hirth}}]{Khater_ActaMater2012}%
  \BibitemOpen
  \bibfield  {author} {\bibinfo {author} {\bibfnamefont {H.~A.}\ \bibnamefont
  {Khater}}, \bibinfo {author} {\bibfnamefont {A.}~\bibnamefont {Serra}},
  \bibinfo {author} {\bibfnamefont {R.~C.}\ \bibnamefont {Pond}}, \ and\
  \bibinfo {author} {\bibfnamefont {J.~P.}\ \bibnamefont {Hirth}},\ }\href@noop
  {} {\bibfield  {journal} {\bibinfo  {journal} {Acta Mater}\ }\textbf
  {\bibinfo {volume} {60}},\ \bibinfo {pages} {2007} (\bibinfo {year}
  {2012})}\BibitemShut {NoStop}%
\bibitem [{\citenamefont {Rajabzadeh}\ \emph {et~al.}(2013)\citenamefont
  {Rajabzadeh}, \citenamefont {Mompiou}, \citenamefont {Legros},\ and\
  \citenamefont {Combe}}]{Rajabzadeh_PRL2013}%
  \BibitemOpen
  \bibfield  {author} {\bibinfo {author} {\bibfnamefont {A.}~\bibnamefont
  {Rajabzadeh}}, \bibinfo {author} {\bibfnamefont {F.}~\bibnamefont {Mompiou}},
  \bibinfo {author} {\bibfnamefont {M.}~\bibnamefont {Legros}}, \ and\ \bibinfo
  {author} {\bibfnamefont {N.}~\bibnamefont {Combe}},\ }\href@noop {}
  {\bibfield  {journal} {\bibinfo  {journal} {Phys. Rev. Lett.}\ }\textbf
  {\bibinfo {volume} {110}},\ \bibinfo {pages} {265507} (\bibinfo {year}
  {2013})}\BibitemShut {NoStop}%
\bibitem [{\citenamefont {Shao}\ \emph {et~al.}(2013)\citenamefont {Shao},
  \citenamefont {Wang}, \citenamefont {Misra},\ and\ \citenamefont
  {Hoagland}}]{Shao_SciRep2013}%
  \BibitemOpen
  \bibfield  {author} {\bibinfo {author} {\bibfnamefont {S.}~\bibnamefont
  {Shao}}, \bibinfo {author} {\bibfnamefont {J.}~\bibnamefont {Wang}}, \bibinfo
  {author} {\bibfnamefont {A.}~\bibnamefont {Misra}}, \ and\ \bibinfo {author}
  {\bibfnamefont {R.~G.}\ \bibnamefont {Hoagland}},\ }\href@noop {} {\bibfield
  {journal} {\bibinfo  {journal} {Sci Rep}\ }\textbf {\bibinfo {volume} {3}},\
  \bibinfo {pages} {2448} (\bibinfo {year} {2013})}\BibitemShut {NoStop}%
\bibitem [{\citenamefont {Beyerlein}\ \emph {et~al.}(2015)\citenamefont
  {Beyerlein}, \citenamefont {Demkowicz}, \citenamefont {Misra},\ and\
  \citenamefont {Uberuaga}}]{Beyerlein_ProgMaterSci2015}%
  \BibitemOpen
  \bibfield  {author} {\bibinfo {author} {\bibfnamefont {I.~J.}\ \bibnamefont
  {Beyerlein}}, \bibinfo {author} {\bibfnamefont {M.~J.}\ \bibnamefont
  {Demkowicz}}, \bibinfo {author} {\bibfnamefont {A.}~\bibnamefont {Misra}}, \
  and\ \bibinfo {author} {\bibfnamefont {B.~P.}\ \bibnamefont {Uberuaga}},\
  }\href@noop {} {\bibfield  {journal} {\bibinfo  {journal} {Prog. Mater.
  Sci.}\ }\textbf {\bibinfo {volume} {74}},\ \bibinfo {pages} {125} (\bibinfo
  {year} {2015})}\BibitemShut {NoStop}%
\bibitem [{\citenamefont {Field}\ \emph {et~al.}(2015)\citenamefont {Field},
  \citenamefont {Yang}, \citenamefont {Allen},\ and\ \citenamefont
  {Busby}}]{Field_ActaMater2015}%
  \BibitemOpen
  \bibfield  {author} {\bibinfo {author} {\bibfnamefont {K.~G.}\ \bibnamefont
  {Field}}, \bibinfo {author} {\bibfnamefont {Y.}~\bibnamefont {Yang}},
  \bibinfo {author} {\bibfnamefont {T.~R.}\ \bibnamefont {Allen}}, \ and\
  \bibinfo {author} {\bibfnamefont {J.~T.}\ \bibnamefont {Busby}},\ }\href@noop
  {} {\bibfield  {journal} {\bibinfo  {journal} {Acta Mater.}\ }\textbf
  {\bibinfo {volume} {89}},\ \bibinfo {pages} {438} (\bibinfo {year}
  {2015})}\BibitemShut {NoStop}%
\bibitem [{\citenamefont {Ardell}\ and\ \citenamefont
  {Bellon}(2016)}]{Ardell_COSSMS2016}%
  \BibitemOpen
  \bibfield  {author} {\bibinfo {author} {\bibfnamefont {A.~J.}\ \bibnamefont
  {Ardell}}\ and\ \bibinfo {author} {\bibfnamefont {P.}~\bibnamefont
  {Bellon}},\ }\href@noop {} {\bibfield  {journal} {\bibinfo  {journal} {Curr.
  Opin. Solid. St. M.}\ }\textbf {\bibinfo {volume} {20}},\ \bibinfo {pages}
  {115} (\bibinfo {year} {2016})}\BibitemShut {NoStop}%
\bibitem [{\citenamefont {Hirth}\ and\ \citenamefont
  {Lothe}(1982)}]{Hirth_book1982}%
  \BibitemOpen
  \bibfield  {author} {\bibinfo {author} {\bibfnamefont {J.~P.}\ \bibnamefont
  {Hirth}}\ and\ \bibinfo {author} {\bibfnamefont {J.}~\bibnamefont {Lothe}},\
  }\href@noop {} {\emph {\bibinfo {title} {Theory of dislocations}}}\ (\bibinfo
   {publisher} {Wiley},\ \bibinfo {address} {New York},\ \bibinfo {year}
  {1982})\BibitemShut {NoStop}%
\bibitem [{\citenamefont {Arzt}\ \emph {et~al.}(1983)\citenamefont {Arzt},
  \citenamefont {Ashby},\ and\ \citenamefont {Verrall}}]{Arzt_ActaMetall1983}%
  \BibitemOpen
  \bibfield  {author} {\bibinfo {author} {\bibfnamefont {E.}~\bibnamefont
  {Arzt}}, \bibinfo {author} {\bibfnamefont {M.~F.}\ \bibnamefont {Ashby}}, \
  and\ \bibinfo {author} {\bibfnamefont {R.~A.}\ \bibnamefont {Verrall}},\
  }\href@noop {} {\bibfield  {journal} {\bibinfo  {journal} {Acta Metall}\
  }\textbf {\bibinfo {volume} {31}},\ \bibinfo {pages} {1977 } (\bibinfo {year}
  {1983})}\BibitemShut {NoStop}%
\bibitem [{\citenamefont {Bachurin}\ \emph {et~al.}(2012)\citenamefont
  {Bachurin}, \citenamefont {Nazarov},\ and\ \citenamefont
  {Weissm眉ller}}]{Bachurin_ActaMater2012}%
  \BibitemOpen
  \bibfield  {author} {\bibinfo {author} {\bibfnamefont {D.~V.}\ \bibnamefont
  {Bachurin}}, \bibinfo {author} {\bibfnamefont {A.~A.}\ \bibnamefont
  {Nazarov}}, \ and\ \bibinfo {author} {\bibfnamefont {J.}~\bibnamefont
  {Weissm眉ller}},\ }\href@noop {} {\bibfield  {journal} {\bibinfo  {journal}
  {Acta Mater.}\ }\textbf {\bibinfo {volume} {60}},\ \bibinfo {pages} {7064}
  (\bibinfo {year} {2012})}\BibitemShut {NoStop}%
\bibitem [{\citenamefont {Gu}\ \emph {et~al.}(2017)\citenamefont {Gu},
  \citenamefont {Han}, \citenamefont {Dai}, \citenamefont {Zhu}, \citenamefont
  {Xiang},\ and\ \citenamefont {Srolovitz}}]{GuYJ_JMPS2017}%
  \BibitemOpen
  \bibfield  {author} {\bibinfo {author} {\bibfnamefont {Y.~J.}\ \bibnamefont
  {Gu}}, \bibinfo {author} {\bibfnamefont {J.}~\bibnamefont {Han}}, \bibinfo
  {author} {\bibfnamefont {S.~Y.}\ \bibnamefont {Dai}}, \bibinfo {author}
  {\bibfnamefont {Y.~C.}\ \bibnamefont {Zhu}}, \bibinfo {author} {\bibfnamefont
  {Y.}~\bibnamefont {Xiang}}, \ and\ \bibinfo {author} {\bibfnamefont {D.~J.}\
  \bibnamefont {Srolovitz}},\ }\href@noop {} {\bibfield  {journal} {\bibinfo
  {journal} {J. Mech. Phys. Solids}\ }\textbf {\bibinfo {volume} {101}},\
  \bibinfo {pages} {166} (\bibinfo {year} {2017})}\BibitemShut {NoStop}%
\bibitem [{\citenamefont {Gu}\ \emph {et~al.}(2015)\citenamefont {Gu},
  \citenamefont {Xiang}, \citenamefont {Quek},\ and\ \citenamefont
  {Srolovitz}}]{GuYJ_JMPS2015}%
  \BibitemOpen
  \bibfield  {author} {\bibinfo {author} {\bibfnamefont {Y.~J.}\ \bibnamefont
  {Gu}}, \bibinfo {author} {\bibfnamefont {Y.}~\bibnamefont {Xiang}}, \bibinfo
  {author} {\bibfnamefont {S.~S.}\ \bibnamefont {Quek}}, \ and\ \bibinfo
  {author} {\bibfnamefont {D.~J.}\ \bibnamefont {Srolovitz}},\ }\href@noop {}
  {\bibfield  {journal} {\bibinfo  {journal} {J. Mech. Phys. Solids}\ }\textbf
  {\bibinfo {volume} {83}},\ \bibinfo {pages} {319} (\bibinfo {year}
  {2015})}\BibitemShut {NoStop}%
\bibitem [{\citenamefont {Ham}(1959)}]{Ham_JAP1959}%
  \BibitemOpen
  \bibfield  {author} {\bibinfo {author} {\bibfnamefont {F.~S.}\ \bibnamefont
  {Ham}},\ }\href@noop {} {\bibfield  {journal} {\bibinfo  {journal} {J. Appl.
  Phys.}\ }\textbf {\bibinfo {volume} {30}},\ \bibinfo {pages} {915} (\bibinfo
  {year} {1959})}\BibitemShut {NoStop}%
\bibitem [{\citenamefont {Geslin}\ \emph {et~al.}(2015)\citenamefont {Geslin},
  \citenamefont {Appolaire},\ and\ \citenamefont {Finel}}]{Geslin_PRL2015}%
  \BibitemOpen
  \bibfield  {author} {\bibinfo {author} {\bibfnamefont {P.~A.}\ \bibnamefont
  {Geslin}}, \bibinfo {author} {\bibfnamefont {B.}~\bibnamefont {Appolaire}}, \
  and\ \bibinfo {author} {\bibfnamefont {A.}~\bibnamefont {Finel}},\
  }\href@noop {} {\bibfield  {journal} {\bibinfo  {journal} {Phys. Rev. Lett.}\
  }\textbf {\bibinfo {volume} {115}},\ \bibinfo {pages} {265501} (\bibinfo
  {year} {2015})}\BibitemShut {NoStop}%
\bibitem [{\citenamefont {Chapman}\ \emph {et~al.}(2016)\citenamefont
  {Chapman}, \citenamefont {Xiang},\ and\ \citenamefont
  {Zhu}}]{Chapman_SIAP2016}%
  \BibitemOpen
  \bibfield  {author} {\bibinfo {author} {\bibfnamefont {S.~J.}\ \bibnamefont
  {Chapman}}, \bibinfo {author} {\bibfnamefont {Y.}~\bibnamefont {Xiang}}, \
  and\ \bibinfo {author} {\bibfnamefont {Y.~C.}\ \bibnamefont {Zhu}},\
  }\href@noop {} {\bibfield  {journal} {\bibinfo  {journal} {SIAM J. Appl.
  Math.}\ }\textbf {\bibinfo {volume} {76}},\ \bibinfo {pages} {750} (\bibinfo
  {year} {2016})}\BibitemShut {NoStop}%
\bibitem [{\citenamefont {Zhu}\ \emph {et~al.}(2016)\citenamefont {Zhu},
  \citenamefont {Niu},\ and\ \citenamefont {Xiang}}]{ZhuYC_JMPS2016}%
  \BibitemOpen
  \bibfield  {author} {\bibinfo {author} {\bibfnamefont {Y.~C.}\ \bibnamefont
  {Zhu}}, \bibinfo {author} {\bibfnamefont {X.~H.}\ \bibnamefont {Niu}}, \ and\
  \bibinfo {author} {\bibfnamefont {Y.}~\bibnamefont {Xiang}},\ }\href@noop {}
  {\bibfield  {journal} {\bibinfo  {journal} {J. Mech. Phys. Solids}\ }\textbf
  {\bibinfo {volume} {96}},\ \bibinfo {pages} {369} (\bibinfo {year}
  {2016})}\BibitemShut {NoStop}%
\bibitem [{Sup()}]{Supp_PRL2018}%
  \BibitemOpen
  \href@noop {} {\enquote {\bibinfo {title} {Supplemental material},}\
  }\BibitemShut {NoStop}%
\bibitem [{\citenamefont {Chapman}\ \emph {et~al.}(2015)\citenamefont
  {Chapman}, \citenamefont {Hewett},\ and\ \citenamefont
  {Trefethen}}]{Chapman_SIAMRev2015}%
  \BibitemOpen
  \bibfield  {author} {\bibinfo {author} {\bibfnamefont {S.~J.}\ \bibnamefont
  {Chapman}}, \bibinfo {author} {\bibfnamefont {D.}~\bibnamefont {Hewett}}, \
  and\ \bibinfo {author} {\bibfnamefont {L.}~\bibnamefont {Trefethen}},\
  }\href@noop {} {\bibfield  {journal} {\bibinfo  {journal} {SIAM Rev.}\
  }\textbf {\bibinfo {volume} {57}},\ \bibinfo {pages} {398} (\bibinfo {year}
  {2015})}\BibitemShut {NoStop}%
\bibitem [{\citenamefont {Kang}\ \emph {et~al.}(2012)\citenamefont {Kang},
  \citenamefont {Wang},\ and\ \citenamefont {Beyerlein}}]{KangK_JAP2012}%
  \BibitemOpen
  \bibfield  {author} {\bibinfo {author} {\bibfnamefont {K.}~\bibnamefont
  {Kang}}, \bibinfo {author} {\bibfnamefont {J.}~\bibnamefont {Wang}}, \ and\
  \bibinfo {author} {\bibfnamefont {I.~J.}\ \bibnamefont {Beyerlein}},\
  }\href@noop {} {\bibfield  {journal} {\bibinfo  {journal} {J. Appl. Phys.}\
  }\textbf {\bibinfo {volume} {111}},\ \bibinfo {pages} {053531} (\bibinfo
  {year} {2012})}\BibitemShut {NoStop}%
\bibitem [{\citenamefont {Hirth}\ \emph {et~al.}(2013)\citenamefont {Hirth},
  \citenamefont {Pond}, \citenamefont {Hoagland}, \citenamefont {Liu},\ and\
  \citenamefont {Wang}}]{Hirth_ProgMaterSci2013}%
  \BibitemOpen
  \bibfield  {author} {\bibinfo {author} {\bibfnamefont {J.~P.}\ \bibnamefont
  {Hirth}}, \bibinfo {author} {\bibfnamefont {R.~C.}\ \bibnamefont {Pond}},
  \bibinfo {author} {\bibfnamefont {R.~G.}\ \bibnamefont {Hoagland}}, \bibinfo
  {author} {\bibfnamefont {X.~Y.}\ \bibnamefont {Liu}}, \ and\ \bibinfo
  {author} {\bibfnamefont {J.}~\bibnamefont {Wang}},\ }\href@noop {} {\bibfield
   {journal} {\bibinfo  {journal} {Prog. Mater. Sci.}\ }\textbf {\bibinfo
  {volume} {58}},\ \bibinfo {pages} {749} (\bibinfo {year} {2013})}\BibitemShut
  {NoStop}%
\bibitem [{\citenamefont {Wang}\ \emph {et~al.}(2013)\citenamefont {Wang},
  \citenamefont {Zhang}, \citenamefont {Zhou}, \citenamefont {Beyerlein},\ and\
  \citenamefont {Misra}}]{WangJ_JMaterRes2013}%
  \BibitemOpen
  \bibfield  {author} {\bibinfo {author} {\bibfnamefont {J.}~\bibnamefont
  {Wang}}, \bibinfo {author} {\bibfnamefont {R.~F.}\ \bibnamefont {Zhang}},
  \bibinfo {author} {\bibfnamefont {C.~Z.}\ \bibnamefont {Zhou}}, \bibinfo
  {author} {\bibfnamefont {I.~J.}\ \bibnamefont {Beyerlein}}, \ and\ \bibinfo
  {author} {\bibfnamefont {A.}~\bibnamefont {Misra}},\ }\href@noop {}
  {\bibfield  {journal} {\bibinfo  {journal} {J. Mater. Res.}\ }\textbf
  {\bibinfo {volume} {28}},\ \bibinfo {pages} {1646} (\bibinfo {year}
  {2013})}\BibitemShut {NoStop}%
\bibitem [{\citenamefont {Rose}\ \emph {et~al.}(1997)\citenamefont {Rose},
  \citenamefont {Balogh},\ and\ \citenamefont {Hahn}}]{Rose_NIMPRB1997}%
  \BibitemOpen
  \bibfield  {author} {\bibinfo {author} {\bibfnamefont {M.}~\bibnamefont
  {Rose}}, \bibinfo {author} {\bibfnamefont {A.~G.}\ \bibnamefont {Balogh}}, \
  and\ \bibinfo {author} {\bibfnamefont {H.}~\bibnamefont {Hahn}},\ }\href@noop
  {} {\bibfield  {journal} {\bibinfo  {journal} {Nucl. Instrum. Meth. Phys.
  Res. B}\ }\textbf {\bibinfo {volume} {127-128}},\ \bibinfo {pages} {119}
  (\bibinfo {year} {1997})}\BibitemShut {NoStop}%
\bibitem [{\citenamefont {Zhu}\ and\ \citenamefont
  {Xiang}(2014)}]{ZhuXH_JMPS2014}%
  \BibitemOpen
  \bibfield  {author} {\bibinfo {author} {\bibfnamefont {X.~H.}\ \bibnamefont
  {Zhu}}\ and\ \bibinfo {author} {\bibfnamefont {Y.}~\bibnamefont {Xiang}},\
  }\href@noop {} {\bibfield  {journal} {\bibinfo  {journal} {J. Mech. Phys.
  Solids}\ }\textbf {\bibinfo {volume} {69}},\ \bibinfo {pages} {175} (\bibinfo
  {year} {2014})}\BibitemShut {NoStop}%
\bibitem [{\citenamefont {Sutton}\ and\ \citenamefont
  {Balluffi}(1995)}]{Sutton_book1995}%
  \BibitemOpen
  \bibfield  {author} {\bibinfo {author} {\bibfnamefont {A.~P.}\ \bibnamefont
  {Sutton}}\ and\ \bibinfo {author} {\bibfnamefont {R.~W.}\ \bibnamefont
  {Balluffi}},\ }\href@noop {} {\emph {\bibinfo {title} {Interfaces in
  Crystalline Materials}}}\ (\bibinfo  {publisher} {Oxford University Press},\
  \bibinfo {address} {New York},\ \bibinfo {year} {1995})\BibitemShut {NoStop}%
\bibitem [{\citenamefont {Zhang}\ and\ \citenamefont
  {Xiang}(2017)}]{ZhangLC_arXiv2017}%
  \BibitemOpen
  \bibfield  {author} {\bibinfo {author} {\bibfnamefont {L.~C.}\ \bibnamefont
  {Zhang}}\ and\ \bibinfo {author} {\bibfnamefont {Y.}~\bibnamefont {Xiang}},\
  }\href@noop {} {\bibfield  {journal} {\bibinfo  {journal}
  {arxiv.org/1710.01856}\ } (\bibinfo {year} {2017})}\BibitemShut {NoStop}%
\bibitem [{\citenamefont {Keralavarma}\ \emph {et~al.}(2012)\citenamefont
  {Keralavarma}, \citenamefont {Cagin}, \citenamefont {Arsenlis},\ and\
  \citenamefont {Benzerga}}]{Keralavarma_PRL2012}%
  \BibitemOpen
  \bibfield  {author} {\bibinfo {author} {\bibfnamefont {S.~M.}\ \bibnamefont
  {Keralavarma}}, \bibinfo {author} {\bibfnamefont {T.}~\bibnamefont {Cagin}},
  \bibinfo {author} {\bibfnamefont {A.}~\bibnamefont {Arsenlis}}, \ and\
  \bibinfo {author} {\bibfnamefont {A.~A.}\ \bibnamefont {Benzerga}},\
  }\href@noop {} {\bibfield  {journal} {\bibinfo  {journal} {Phys. Rev. Lett.}\
  }\textbf {\bibinfo {volume} {109}} (\bibinfo {year} {2012})}\BibitemShut
  {NoStop}%
\bibitem [{\citenamefont {Niu}\ \emph {et~al.}(2017)\citenamefont {Niu},
  \citenamefont {Luo}, \citenamefont {Lu},\ and\ \citenamefont
  {Xiang}}]{NiuXH_JMPS2017}%
  \BibitemOpen
  \bibfield  {author} {\bibinfo {author} {\bibfnamefont {X.~H.}\ \bibnamefont
  {Niu}}, \bibinfo {author} {\bibfnamefont {T.}~\bibnamefont {Luo}}, \bibinfo
  {author} {\bibfnamefont {J.~F.}\ \bibnamefont {Lu}}, \ and\ \bibinfo {author}
  {\bibfnamefont {Y.}~\bibnamefont {Xiang}},\ }\href@noop {} {\bibfield
  {journal} {\bibinfo  {journal} {J. Mech. Phys. Solids}\ }\textbf {\bibinfo
  {volume} {99}},\ \bibinfo {pages} {242} (\bibinfo {year} {2017})}\BibitemShut
  {NoStop}%
\bibitem [{\citenamefont {Mordehai}\ \emph {et~al.}(2008)\citenamefont
  {Mordehai}, \citenamefont {Clouet}, \citenamefont {Fivel},\ and\
  \citenamefont {Verdier}}]{Mordehai_PhilMag2008}%
  \BibitemOpen
  \bibfield  {author} {\bibinfo {author} {\bibfnamefont {D.}~\bibnamefont
  {Mordehai}}, \bibinfo {author} {\bibfnamefont {E.}~\bibnamefont {Clouet}},
  \bibinfo {author} {\bibfnamefont {M.}~\bibnamefont {Fivel}}, \ and\ \bibinfo
  {author} {\bibfnamefont {M.}~\bibnamefont {Verdier}},\ }\href@noop {}
  {\bibfield  {journal} {\bibinfo  {journal} {Philos. Mag.}\ }\textbf {\bibinfo
  {volume} {88}},\ \bibinfo {pages} {899} (\bibinfo {year} {2008})}\BibitemShut
  {NoStop}%
\bibitem [{\citenamefont {Geers}\ \emph {et~al.}(2014)\citenamefont {Geers},
  \citenamefont {Cottura}, \citenamefont {Appolaire}, \citenamefont {Busso},
  \citenamefont {Forest},\ and\ \citenamefont {Villani}}]{Geers_JMPS2014}%
  \BibitemOpen
  \bibfield  {author} {\bibinfo {author} {\bibfnamefont {M.~G.~D.}\
  \bibnamefont {Geers}}, \bibinfo {author} {\bibfnamefont {M.}~\bibnamefont
  {Cottura}}, \bibinfo {author} {\bibfnamefont {B.}~\bibnamefont {Appolaire}},
  \bibinfo {author} {\bibfnamefont {E.~P.}\ \bibnamefont {Busso}}, \bibinfo
  {author} {\bibfnamefont {S.}~\bibnamefont {Forest}}, \ and\ \bibinfo {author}
  {\bibfnamefont {A.}~\bibnamefont {Villani}},\ }\href@noop {} {\bibfield
  {journal} {\bibinfo  {journal} {J. Mech. Phys. Solids}\ }\textbf {\bibinfo
  {volume} {70}},\ \bibinfo {pages} {136} (\bibinfo {year} {2014})}\BibitemShut
  {NoStop}%
\bibitem [{\citenamefont {Zhu}\ \emph {et~al.}(2017)\citenamefont {Zhu},
  \citenamefont {Wang}, \citenamefont {Xiang},\ and\ \citenamefont
  {Guo}}]{ZhuYC_JMPS2017a}%
  \BibitemOpen
  \bibfield  {author} {\bibinfo {author} {\bibfnamefont {Y.~C.}\ \bibnamefont
  {Zhu}}, \bibinfo {author} {\bibfnamefont {J.}~\bibnamefont {Wang}}, \bibinfo
  {author} {\bibfnamefont {Y.}~\bibnamefont {Xiang}}, \ and\ \bibinfo {author}
  {\bibfnamefont {X.}~\bibnamefont {Guo}},\ }\href@noop {} {\bibfield
  {journal} {\bibinfo  {journal} {J. Mech. Phys. Solids}\ }\textbf {\bibinfo
  {volume} {105}},\ \bibinfo {pages} {1} (\bibinfo {year} {2017})}\BibitemShut
  {NoStop}%
\bibitem [{\citenamefont {Rovelli}\ \emph {et~al.}(2017)\citenamefont
  {Rovelli}, \citenamefont {Dudarev},\ and\ \citenamefont
  {Sutton}}]{Rovelli_JMPS2017}%
  \BibitemOpen
  \bibfield  {author} {\bibinfo {author} {\bibfnamefont {I.}~\bibnamefont
  {Rovelli}}, \bibinfo {author} {\bibfnamefont {S.~L.}\ \bibnamefont
  {Dudarev}}, \ and\ \bibinfo {author} {\bibfnamefont {A.~P.}\ \bibnamefont
  {Sutton}},\ }\href@noop {} {\bibfield  {journal} {\bibinfo  {journal} {J.
  Mech. Phys. Solids}\ }\textbf {\bibinfo {volume} {103}},\ \bibinfo {pages}
  {121} (\bibinfo {year} {2017})}\BibitemShut {NoStop}%
\end{thebibliography}%

\end{document}